\documentclass[aps,prb,twocolumn,showpacs,superscriptaddress]{revtex4}
\usepackage{latexsym}
\usepackage{graphicx}
\usepackage{amssymb}
\usepackage{amsmath}
\begin{document}

\title{Magnetic properties of FeSe superconductor}
\author{G.E. Grechnev}
\email{grechnev@ilt.kharkov.ua}
\affiliation{B. Verkin Institute for Low Temperature Physics and
Engineering, National Academy of Sciences of Ukraine, 61103
Kharkov, Ukraine}

\author{A.S. Panfilov}
\affiliation{B. Verkin Institute for Low Temperature Physics and
Engineering, National Academy of Sciences of Ukraine, 61103
Kharkov, Ukraine}

\author{V.A. Desnenko}
\affiliation{B. Verkin Institute for Low Temperature Physics and
Engineering, National Academy of Sciences of Ukraine, 61103
Kharkov, Ukraine}

\author{A.V. Fedorchenko}
\affiliation{B. Verkin Institute for Low Temperature Physics and
Engineering, National Academy of Sciences of Ukraine, 61103
Kharkov, Ukraine}

\author{S.L. Gnatchenko}
\affiliation{B. Verkin Institute for Low Temperature Physics and
Engineering, National Academy of Sciences of Ukraine, 61103
Kharkov, Ukraine}

\author{D.A. Chareev}
\affiliation {Institute of Experimental Mineralogy, Russian
Academy of Sciences,\\ Chernogolovka, Moscow Region 142432,
Russia}

\author{O.S. Volkova}
\affiliation{Department of Low Temperature Physics, Moscow State University,
Moscow 119991, Russia}

\author{A.N. Vasiliev}
\affiliation{Department of Low Temperature Physics, Moscow State University,
Moscow 119991, Russia}

\begin{abstract}
A detailed magnetization study for the novel FeSe superconductor
is carried out to investigate the behavior of the intrinsic
magnetic susceptibility $\chi$ in the normal state with
temperature and under hydrostatic pressure. The temperature
dependencies of $\chi$ and its anisotropy $\Delta
\chi=\chi_{\|}-\chi_{\bot}$ are measured for FeSe single crystals
in the temperature range $4.2-300$~K, and a substantial growth of
susceptibility with temperature is revealed. The observed
anisotropy $\Delta \chi$ is very large and comparable with the
averaged susceptibility at low temperatures. For a polycrystalline
sample of FeSe, a significant pressure effect on $\chi$ is
determined to be essentially dependent on temperature. {\em Ab
initio} calculations of the pressure dependent electronic
structure and magnetic susceptibility indicate that FeSe is close
to magnetic instability with dominating enhanced spin
paramagnetism. The calculated paramagnetic susceptibility exhibits
a strong dependence on the unit cell volume and especially on the
height $Z$ of chalcogen species from the Fe plane. The change of
$Z$ under pressure determines a large positive pressure effect on
$\chi$ which is observed at low temperatures. It is shown that the
literature experimental data on the strong and nonmonotonic
pressure dependence of the superconducting transition temperature
in FeSe correlate qualitatively with calculated behavior of the
density of electronic states at the Fermi level.
\end{abstract}



\maketitle

\section{Introduction}
Soon after the discovery of superconductivity in
LaFeAsO$_{1-x}$F$_y$, the superconductivity was also detected in
the binary compound FeSe$_{1-x}$ \cite{Hsu} with transition
temperature $T_{\rm c}\simeq 8$\,K.
This compound possesses the simplest crystal structure among the new families of Fe-based
superconductors, and consists of a stack of Fe square planar
layers, which are tetrahedrally coordinated by Se atoms.
Also the large pressure effect on transition temperature was later observed
\cite{Medvedev,Margadonna,Braithwaite} with $T_{\rm c}\approx 37$\,K
at pressures $P\approx9$\,GPa, indicating that FeSe$_{1-x}$
is actually a high temperature superconductor.
Therefore, the superconducting FeSe$_{1-x}$ compound has attracted considerable
attention and is a subject of intensive studies for the last years
\cite{Mcqueen,Pomjakushina,Millican,Mizuguchi1,Hu,Lin,Fedorchenko,Mizuguchi2}.
The structural simplicity of FeSe favors experimental and theoretical studies of
chemical substitution and high pressure effects, which are aimed
at promoting a better understanding of a mechanism of the
superconductivity.

Upon cooling below room temperature, the tetragonal $P4/nmm$ phase
of FeSe$_{1-x}$ undergoes a subtle distortion to the lower
symmetry orthorhombic $Cmma$ phase
\cite{Margadonna,Mcqueen,Pomjakushina,Millican}. This transition
occurs within a broad temperature range, about 70$\div$100 K,
depending on the stoichiometry of the samples. It was also found
that the tetragonal phase undergoes structural transitions under
high pressures ($P\ge 10$ GPa) to the hexagonal
non-superconducting $P6_{3}mmc$ NiAs-type phase
\cite{Medvedev,Margadonna,Mcqueen}, and then to its orthorhombic
modification ($Pbnm$, MnP-type) \cite{Medvedev,Garbarino,Kumar}.
The recent theoretical examination of stability regions of the
high pressure phases of FeSe$_{1-x}$ \cite{Chadov,Freeman}
indicated a possibility of metallization and superconductivity in
the orthorhombic $Pbnm$ phase under high pressure.

Though a substantial increase of $T_{\rm c}$ was clearly observed
under pressure \cite{Medvedev,Margadonna,Masaki,Miyoshi,Okabe},
these studies did not detect any magnetic ordering. However,
recent NMR studies provided some indication of an incipient
magnetic phase transition under pressure \cite{Imai}. Recently, a
static magnetic ordering has been detected above $P\sim 1$\,GPa by
means of zero-field muon spin rotation (ZF $\mu$SR)
\cite{BendelePRL,Bendele} and neutron diffraction \cite{Bendele}.
These studies have revealed that as soon as magnetic ordering
emerges, the magnetic and superconducting states apparently
coexist, and both the magnetic ordering temperature $T_{\rm N}$
and $T_{\rm c}$ increase simultaneously with increasing pressure.
Also, it was recently found, that upon applying pressure the
increase of $T_{\rm c}$ in FeSe$_{1-x}$ appeared to be
nonmonotonic and exhibits a local maximum at $P\simeq$ 0.8 GPa,
which is followed by a local minimum at $P\sim$ 1.2 GPa
\cite{Masaki,Miyoshi,BendelePRL,Bendele} . Therefore, there is
still a considerable controversy regarding an interplay between
electronic structure, magnetism and superconductivity in the
FeSe$_{1-x}$ compound, especially under pressure.

The experimental data on magnetic susceptibility of FeSe$_{1-x}$
system in the normal state are incomplete and contradicting
\cite{Mizuguchi1,Mizuguchi2}. Also, these data are mostly obtained
on polycrystalline samples and often distorted by the presence of
secondary magnetic phases of iron. In order to shed more light on
the relation between magnetic and superconducting properties, it
is very important to elucidate the intrinsic susceptibility of
FeSe$_{1-x}$ superconductors and investigate its evolution with
doping, temperature, and pressure. Here we report on results of
the experimental studies of magnetic properties for
single-crystalline and polycrystalline FeSe samples of high
quality in the normal state.
These studies include measurements of the
temperature dependence of magnetic susceptibility and its
anisotropy as well as the hydrostatic pressure effects.
The experimental investigations are supplemented by {\em ab initio}
calculations of the electronic structure and magnetic
susceptibility of FeSe.
The calculations are based on the local-spin-density approximation
(LSDA) of the density-functional theory (DFT).
The results of experiments and calculations are used
to analyze the nature of magnetism in FeSe and the basic
mechanisms of its strong pressure dependence.

\section{Experimental details and results}

The plate-like single crystals of FeSe$_{1-x}$ superconductor were
grown in evacuated quartz ampoules using the KCl/AlCl$_3$ flux
technique with a constant temperature gradient of 5$^{\circ}$ C per cm
along the ampoule length (temperature of the hot end was kept at
427$^{\circ}$ C, temperature of the cold end was about 380$^{\circ}$ C).
Typical dimensions of the produced single crystalline samples are
(2$\div$3)$\times$(2$\div$3)$\times$(0.3$\div$0.5) mm$^3$.
The tetragonal $P4/nmm$ structure was demonstrated at room temperature
by X-ray diffraction technique.
The energy dispersive X-ray
spectroscopy, performed on a CAMECA SX100 (15 keV) analytical
scanning electron microscope, revealed Fe:Se=1:0.96$\pm$0.02
composition denoted in the following as FeSe for simplicity.

\begin{figure}[]
\begin{center}
\includegraphics*[trim=0mm 0mm 0mm 0mm,scale=0.9]{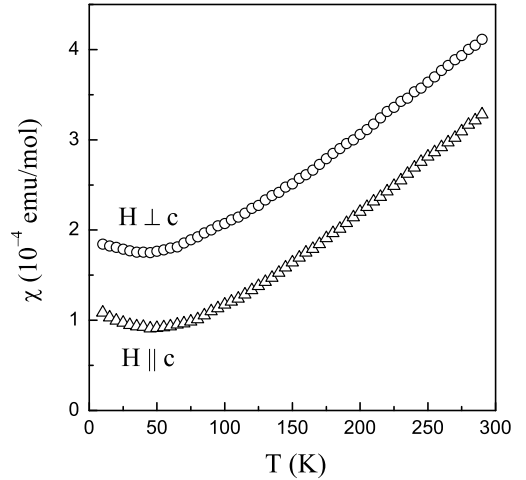}
\end{center}
\caption{\label{X(T)}Temperature dependence of the magnetic
susceptibility in the normal state for the single-crystalline FeSe
measured in the field $H=50$ mT. Data corresponding to the
magnetic field directions $H\perp c$ axis and $H\parallel c$ are
denoted by {\Large$\circ$} and $\bigtriangleup$ symbols,
respectively.}
\end{figure}

The study of magnetic properties of FeSe samples at ambient
pressure was carried out at $T=4.2\div 300$ K by using a SQUID
magnetometer. The superconducting transition was detected within
6$\div$8 K. The magnetization dependencies $M(H)$ in magnetic
fields up to 5 T appeared to be close to linear, indicating that
the concentrations of ferromagnetic impurities are negligibly
small. A typical temperature dependence of the magnetic
susceptibility for single crystalline FeSe samples is shown in
Fig. \ref{X(T)}. As is seen, a substantial growth of
susceptibility with temperature was revealed in the normal state,
as well as a large magnetic anisotropy.

\begin{figure}[b]
\begin{center}
\includegraphics*[trim=0mm 0mm 0mm 0mm,scale=0.9]{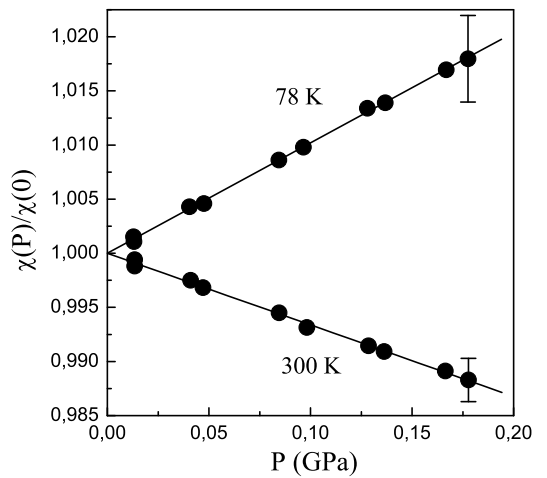}
\end{center}
\caption{\label{X(P)}Pressure dependencies of the magnetic
susceptibility, normalized to its value at $P=0$, for the
polycrystalline FeSe compound at temperatures 78 and 300 K. The
solid lines are guides for the eye.}
\end{figure}

The study of magnetic susceptibility of FeSe under helium gas
pressure $P$ up to 0.2 GPa was performed at fixed temperatures 78
and 300 K by using a pendulum-type magnetometer placed directly in
the nonmagnetic pressure cell \cite{Panfilov}.
In order to measure
the pressure effect with a reasonable accuracy, a sufficiently
large mass of the sample is required.
We have used the FeSe
sample, further called as "polycrystalline" FeSe, which was
prepared by compacting a number of about 50 of small arbitrarily
oriented single crystals inside of an aluminum foil cylinder.
The total mass of the sample was about 200 mg.
The measurements were carried out in the field $H=1.7$ T
and their relative errors did not exceed 0.5\%.
The experimental pressure dependencies $\chi(P)$
at different temperatures are shown in Fig. \ref{X(P)}, which
demonstrates their linear character.
The negative value of the pressure effect,
d\,ln$\chi$/d$P\simeq -6.5\times 10^{-2}$~GPa$^{-1}$,
was observed at temperature 300 K, whereas at
$T=78$ K the effect appeared to be positive,
d\,ln$\chi$/d$P\simeq 10\times 10^{-2}$~GPa$^{-1}$.
The available experimental and
theoretical results on d\,ln$\chi$/d$P$ for FeSe are compiled in
Table \ref{dlnX/dP} together with corresponding data for the
relative FeTe compound for comparison.

\begin{table}[]
\caption{\label{dlnX/dP} Pressure derivatives of the magnetic
susceptibility d\,ln$\chi$/d$P$ (in units 10$^{-2}$ GPa$^{-1}$)
for FeSe and FeTe compounds at different temperatures. }
\vspace{3pt}
\begin{center}
\begin{tabular}{cccc}
\hline\hline & $T$(K) & \multicolumn{2}{c}{d\,ln$\chi$/d$P$} \\
\hline &       & FeSe & FeTe~$^{\rm b}$ \\
 \vspace{-0pt}
 experiment: & 300 & $-6.5\pm 1$ & $13 \pm1$ \\
 & & $\sim -7~^{\rm a}$ & \\
 \vspace{-0pt}
 & 78 &     $10\pm 3$   & $23 \pm 1.5$~ \\
 &    & $\sim 6.5~^{\rm a}$& \\
 & $20$   & $\sim 9~^{\rm a}$& \\
\hline theory: & $ 0 $ & $\simeq 8$ & $\sim 20$  \\ \hline\hline
\end{tabular}
\end{center}\vspace{3pt}
$^{\rm a}$ From NMR Knight shift data of Ref. \onlinecite{Imai}\\
$^{\rm b}$ Results for FeTe are taken from Ref.
\onlinecite{Grechnev0}
\end{table}

\section{Computational details and results}

In order to analyze the magnetic properties of FeSe compound in
the normal state, the {\em ab initio} calculations of the
electronic structure and paramagnetic susceptibility were carried out.
At ambient conditions FeSe compound possesses the tetragonal
PbO-type crystal structure (space group $P4/nmm$), which is
composed by alternating triple-layer slabs.
Each iron layer is sandwiched between two nearest-neighbor layers of Se,
which form edge-shared tetrahedra around the iron sites.
The positions of selenium layers are fixed by the structural parameter $Z$,
which represents the relative height of Se atoms above the iron plane.
The structural parameters of FeSe were determined by means of
X-ray and neutron diffraction
\cite{Margadonna,Pomjakushina,Millican,Hu,Kumar,Gomez}.

The main purpose of the present {\em ab initio} calculations was to
evaluate the paramagnetic response in an external magnetic field and
to elucidate the nature and features of magnetism in the normal state
of FeSe compound.
In the context of this task, the dependencies of the magnetic susceptibility
on volume, lattice parameters and temperature were addressed.
{\em Ab initio} calculations of the electronic structure of FeSe
were performed by employing a full-potential all-electron relativistic
linear muffin-tin orbital method (FP-LMTO, code RSPt~\cite{Wills,Grechnev1}).
The exchange-correlation potential was treated within
the local spin density approximation \cite{Barth} of the DFT.
The calculated basic features of electronic structure of FeSe
appeared to be in a qualitative agreement with results of
previous DFT calculations \cite{Subedi,Pickett,Singh,Ciechan}.
\begin{figure}[]
\begin{center}
\includegraphics[scale=0.42]{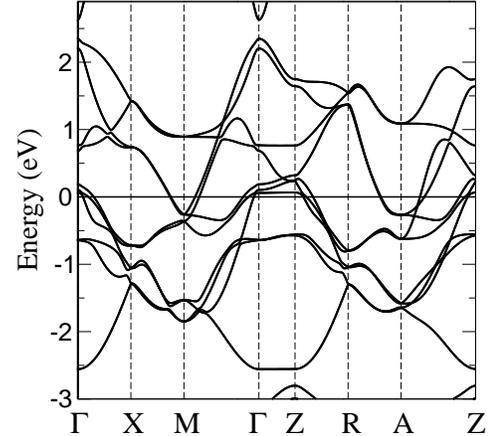}
\caption{\label{fese_bs} Band structure of FeSe around the Fermi
level (at 0~eV) marked by a horizontal line. }
\end{center}
\end{figure}

\begin{figure}[ht]
\begin{center}
\includegraphics*[scale=0.42]{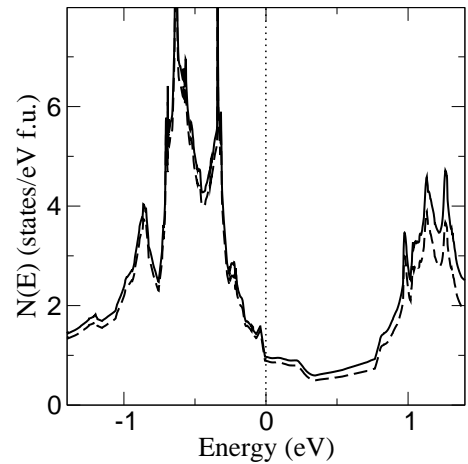}
\caption{\label{fese_dos} Total density of electronic states of
FeSe (solid line) and the partial contribution of the iron
$d$-states (dashed line). The Fermi level position at 0~eV is
marked by a vertical line. }
\end{center}
\end{figure}

As is seen in Figs. \ref{fese_bs} and  \ref{fese_dos}, the
calculated band structure and density of electronic states (DOS)
of FeSe indicate the presence of the hybridized predominantly
$d$-like Fe electronic states close to the Fermi level $E_{\rm F}$.
The chalcogen $p$-states are situated well below $E_{\rm F}$
and slightly hybridized with the $d$-states of iron.
Also, one can see a van Hove singularity in Fig. \ref{fese_dos}
at $\sim40$ meV below $E_{\rm F}$.
It should be noted, that a proximity of the van
Hove singularity to the Fermi level is considered as the key
ingredient for superconductivity in iron-based superconductors
\cite{Kordyuk1}.

\begin{figure}[]
\begin{center}
\includegraphics*[trim=0mm 0mm 0mm 0mm,scale=0.42]{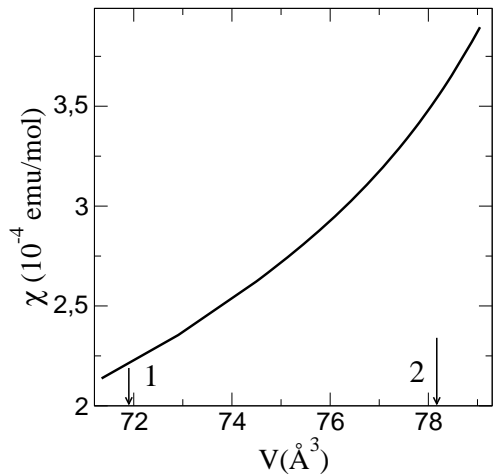}
\caption{\label{X(V)} Calculated paramagnetic susceptibility of
FeSe as a function of the unit cell volume (in \AA$^3$). $Z$ is
taken to be 0.26. The arrows indicate the theoretical (1) and
experimental (2) equilibrium volume values. }
\end{center}
\end{figure}

\begin{figure}[]
\begin{center}
\includegraphics*[trim=0mm 0mm 0mm 0mm,scale=0.42]{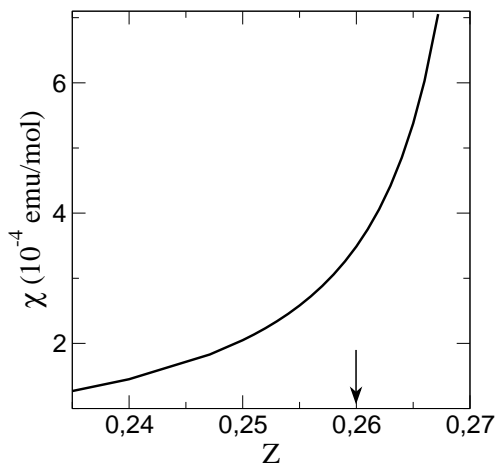}
\caption{\label{X(Z)}
Calculated paramagnetic susceptibility of FeSe as a function of
internal parameter $Z$ for the experimental unit cell volume.
The arrow indicates the experimental value of $Z$.
}
\end{center}
\end{figure}

To evaluate the paramagnetic susceptibility of FeSe,
the FP-LMTO-LSDA calculations of field-induced spin and orbital
(Van Vleck) magnetic moments were carried out within the approach
described in Ref. \onlinecite{Grechnev2}.
The relativistic effects, including spin-orbit coupling, were incorporated,
and the effect of an external magnetic field $\textbf{H}$ was taken into account
self-consistently by means of the Zeeman term:
\begin{equation}
\label{zeeman} {\cal H}_{\rm Z}=\mu_{\rm B} \mathbf{H}\cdot
(2\hat{\bf{s}}+\hat{\bf{l}}) .
\end{equation}
Here $\mu_{\rm B}$ is the Bohr magneton, $\hat{\bf{s}}$ and
$\hat{\bf{l}}$ are the spin and orbital angular momentum
operators, respectively.
The field induced spin and orbital magnetic moments provide
estimations of the related contributions to the magnetic susceptibility,
$\chi_{\rm spin}$ and $\chi_{\rm orb}$.
For the tetragonal crystal structure the components of
these contributions, ${\chi_i}_{\parallel}$ and
${\chi_i}_{\perp}$, are derived from the magnetic moments
calculated in the external field of 10 T, which was applied
parallel and perpendicular to the $c$ axis, respectively.

It is found that magnetic response to the external field is very sensitive
to the unit cell volume, as well as to the structural parameter $Z$,
which represents the relative height of chalcogen species from the Fe plane.
The calculated dependencies of susceptibility of FeSe
as functions of the volume and parameter $Z$ are given in Figs. \ref{X(V)}
and \ref{X(Z)}, respectively.

\begin{figure}[]
\begin{center}
\includegraphics*[trim=0mm 0mm 0mm 0mm,scale=0.42]{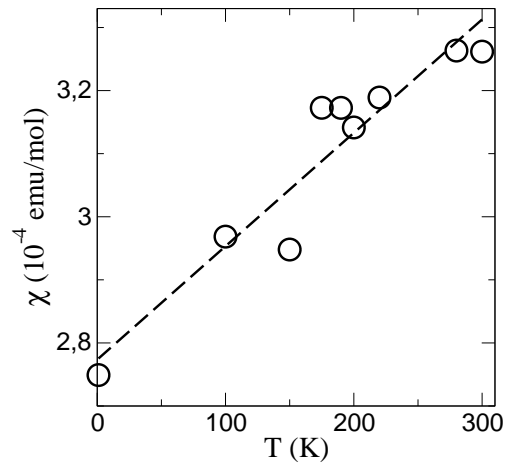}
\caption{\label{XTT}
Calculated temperature dependence of the paramagnetic susceptibility of FeSe.
$Z$ is taken to be 0.26, the unit cell volume is fixed
between the theoretical and experimental values.
The dashed line is a guide for the eye.
}
\end{center}
\end{figure}

Also, the thermal effects are taken into account in order to calculate
the temperature dependence of paramagnetic susceptibility for FeSe compound.
In this case the field induced spin and orbital magnetic moments
were evaluated by corresponding integration with the energy derivative
of the Fermi-Dirac distribution function $f(E,\mu,T)$, and the temperature
dependence of $\chi$ was actually determined by taking into account the finite
width of $-{\rm d}f/{\rm d}E$ (see Ref. \onlinecite{Grechnev1} for details).
It should be noted that the energy derivative of the Fermi-Dirac
distribution corresponds to a Dirac delta function at low temperatures,
having a sharp positive peak at the Fermi energy $E_{\rm F}$.
This steep behavior of $-{\rm d}f/{\rm d}E$ resulted in some instability
in the numerical calculations of $\chi$, which are seen in Fig. \ref{XTT}.

\section{Discussion}

The experimental superconducting transition temperatures 6$\div$8 K,
obtained for the studied FeSe$_{1-x}$ samples, agree with those reported
in literature \cite{Hsu,Margadonna,Braithwaite,Mcqueen,Pomjakushina}.
Above $T_{\rm c}$, a substantial growth of susceptibility with
temperature is revealed in the normal state of FeSe up to 300 K
(Fig.~\ref{X(T)}), what indicates the itinerant nature
of electronic states of Fe at the Fermi energy.

The total susceptibility in the absence of spontaneous magnetic ordering
can be represented as the sum:
\begin{equation}
\label{chitot} \chi_{\rm tot} =\chi_{\rm spin} +\chi_{\rm orb}
+\chi_{\rm dia} +\chi_{\rm L} ,
\end{equation}
where these terms correspond to the Pauli spin susceptibility,
a generalization of the Van Vleck orbital paramagnetism,
the Langevin diamagnetism of closed ion shells, and a generalization
of Landau conduction electrons diamagnetism, respectively.
Obviously, the $\chi_{\rm dia}$ term does not provide any anisotropy,
and for FeSe the Langevin diamagnetism of closed ion shells can be estimated according
to Ref. \onlinecite{Selwood} as $\chi_{\rm dia}\simeq -0.2\times 10^{-4}$ emu/mol.

In order to analyse the experimental data on $\chi $ for FeSe we
used the calculated paramagnetic contributions to susceptibility,
$\chi_{\rm spin}$ and $\chi_{\rm orb}$.
It has been shown \cite{Grechnev2}
that for paramagnetic metallic systems the Stoner
approach underestimates the spin susceptibility, whereas the LSDA
field-induced calculations take into account non-uniform induced
magnetization density in the unit cell and provide more adequate description.
The first-principles calculations of the paramagnetic
susceptibility for FeSe revealed that this system is in close
proximity to the magnetic critical point.
This can be seen from a steep rise of $\chi (V)$ and $\chi (Z)$
in Figs. \ref{X(V)} and \ref{X(Z)}, respectively,
above the corresponding experimental values of $V$ and $Z$.
In fact, the calculated Stoner enhancement
$S\sim 7$ clearly indicates that FeSe is close to the
ferromagnetic Stoner instability.

For the unit cell volume chosen between the theoretical and
experimental values, the dominant spin contribution to magnetic
susceptibility of FeSe is estimated to be
$\chi_{\rm spin}\simeq 2.4\times 10^{-4}$ emu/mol.
The averaged orbital $\chi_{\rm orb}$
term amounts to $\sim 0.4\times 10^{-4}$ emu/mol), being of about
15\% of the total paramagnetism.
From comparison of the calculated paramagnetic susceptibility for
the ground state in Fig.~\ref{XTT},
$\chi_{\rm para}=\chi_{\rm spin}+\chi_{\rm orb}\simeq 2.8\times 10^{-4}$
emu/mol, with the experimental value $\chi_{\rm exp}$ at $T\to 0$ K in
Fig.~\ref{X(T)},
$\chi_{\rm exp}=(\chi_{\parallel}+2\chi_{\perp})/3\simeq 1.5\times 10^{-4}$
emu/mol, it is clear that calculated value  $\chi_{\rm para}$ has
to be substantially compensated by a diamagnetic contribution in
order to conform with the experimental data.
One can estimate the expected diamagnetic contribution
to magnetic susceptibility of FeSe to be about
$(\chi_{\rm exp}-\chi_{\rm para})\sim -1.3\times10^{-4}$ emu/mol.
This diamagnetism is comparable in
absolute value with the paramagnetic contribution, being much
larger than the estimated above Langevin diamagnetism $\chi_{\rm dia}$.
Apparently, it can be ascribed to the $\chi_{\rm L}$ term in Eq. (\ref{chitot}).

According to the experimental data in Fig. \ref{X(T)}, the observed anisotropy of
susceptibility $\Delta \chi$ is large in FeSe,
and even comparable with the averaged susceptibility itself at low temperatures.
It appears to be much larger than the calculated anisotropy of orbital contribution
to paramagnetic susceptibility of FeSe,
$\Delta \chi_{\rm orb} = {\chi_{\rm orb}}_{\parallel}-{\chi_{\rm orb}}_{\perp}
\simeq -0.1\times 10^{-4}$ emu/mol.
Therefore, in order to explain the experimental $\Delta \chi$ one can assume the presence of
a substantial and presumably anisotropic diamagnetic contribution from conduction electrons.

To calculate the Landau diamagnetic contribution $\chi_{\rm L}$ is
a rather difficult problem \cite{Grechnev3,Mikitik}.
The free-electron Landau approximation, which is often used for
estimations, provides $\chi^{0}_{\rm L}$ that equals
$-\frac{1}{3}$ of the Pauli spin susceptibility.
However, for many metallic systems the diamagnetism of conduction electrons
$\chi_{\rm L}$ can be many times larger than the free-electron
Landau estimate $\chi_{\rm L}^{0}$, and such anomalous and
anisotropic diamagnetism is often determined by the presence of
quasi-degenerated states with small effective masses at $E_{\rm F}$
(see Ref. \onlinecite{Mikitik} and references therein).
As can be seen in Fig. \ref{fese_bs}, in FeSe the quasi-degenerate states
with small effective masses exist at $E_{\rm F}$ around the
symmetry points $\Gamma$ and Z, where the band degeneracies are
lifted by the spin-orbital coupling.
Such band structure features
are of particular importance in connection with a manifestation of
the anomalously large and anisotropic $\chi_{\rm L}$, which was
found to originate from the similar degeneracy points \cite{Mikitik}.
It should be emphasized that rigorous theoretical
analysis of $\chi_{\rm L}$ is a rather cumbersome procedure, which
goes beyond the aims of the present work.
At this stage, we have identified appropriate electronic states near $E_{\rm F}$ as
possible sources of the large and anisotropic conduction electrons
diamagnetism in FeSe.

The theoretically evaluated temperature dependence of the
paramagnetic susceptibility $\chi_{\rm para}(T)$ in
Fig.~\ref{XTT}, which takes into account the finite width of the
energy derivative of the Fermi-Dirac distribution function,
actually provides only a slight increase in $\chi$ with temperature.
Thus, the observed substantial growth of $\chi (T)$
is rather puzzling at present.
It is presumably related to a fine structure of DOS
at $E_{\rm F}$, but one should expect that FeSe system is driven
far from the ground state at room temperatures.
At this stage we should admit that increase of the unit cell volume for the
tetragonal $P4/nmm$ phase ($\Delta V/V\simeq 1\%$ with temperature
rising up to 300 K, see Ref. \onlinecite{Mizuguchi2}) can provide
about 10\% growth of paramagnetic susceptibility,
according to the $\chi (V)$ dependence in Fig. \ref{X(V)}.
However, this volume expansion does not explain the
experimental $\chi (T)$ in Fig. \ref{X(T)}.
Also, a change of the
structural parameter $Z$ with temperature can be substantial and
of importance due to the strong $\chi (Z)$ dependence in Fig.
\ref{X(Z)}, but to date the influence of temperature on $Z$ was
not studied in a systematic way.
Therefore, noticeable temperature effects on the lattice
parameters, chalcogen atom position $Z$, and electronic structure
itself should be taken into account in a rigorous quantitative
analysis of $\chi (T)$ in FeSe.

The measured pressure effects on magnetic susceptibility of FeSe
are intriguing and require a detailed examination.
Firstly, as can be seen in Table \ref{dlnX/dP}, there is a striking sign
difference for the pressure effects on $\chi$ at low and room
temperatures.
Also, the absolute value of this effect is substantially larger than that
observed in strongly enhanced itinerant paramagnets \cite{Grechnev2},
and appeared to be comparable with such pressure effect on $\chi$ reported
for the related FeTe compound \cite{Grechnev0} (see  Table \ref{dlnX/dP}).

\begin{figure}[]
\begin{center}
\includegraphics*[trim=0mm 0mm 0mm 0mm,scale=0.89]{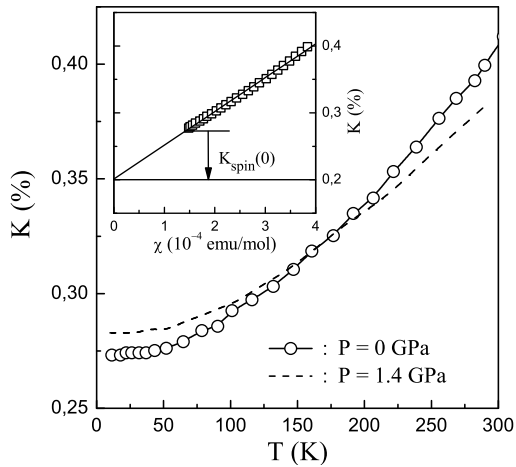}
\end{center}
\caption{\label{NMR}Temperature dependencies of the NMR Knight
shift $K$ in FeSe measured at ambient pressure ({\Large$\circ$}) and at
$P=1.4$ GPa (dashed line). The data are taken from Ref.
\onlinecite{Imai}. The inset shows dependence of $K$ on the
averaged magnetic susceptibility for FeSe,
$\chi=(\chi_{\parallel}+2\chi_{\perp})/3$, from Fig. \ref{X(T)}.}
\end{figure}

It should be noted that the present experimental data on
$\chi(T,P)$ for FeSe are in reasonable agreement with the results
of Ref. \onlinecite{Imai} on temperature and pressure dependencies
of the NMR Knight shift $K$ of FeSe in the normal state.
As can be seen in Fig. \ref{NMR}, the temperature dependence $K(T)$
at ambient pressure reflects the corresponding dependence of the
magnetic susceptibility in Fig. \ref{X(T)}.
Assuming the latter to
be governed by the spin susceptibility $\chi_{\rm spin}(T)$, the
only temperature dependent contribution in $K(T)$ can be
determined as $K_{\rm spin}(T)=\alpha\chi_{\rm spin}(T)$ with
$\alpha\simeq 5\times 10^2$ \%(emu/mol)$^{-1}$, resulted from the
slope of $K~vs~\chi$ linear dependence in inset of Fig. \ref{NMR}.
In addition, this contribution is expected to be also responsible
for the pressure effect on $K$.
Using a rough approximation
$\chi_{\rm spin}(T)\approx\chi(T)$ we obtain an estimate of
d\,ln$\chi$/d$P\approx$ d\,ln$K_{\rm spin}$/d$P$, which is listed
in Table \ref{dlnX/dP}.

\begin{figure}[b]
\begin{center}
\includegraphics*[trim=0mm 0mm 0mm 0mm,scale=0.89]{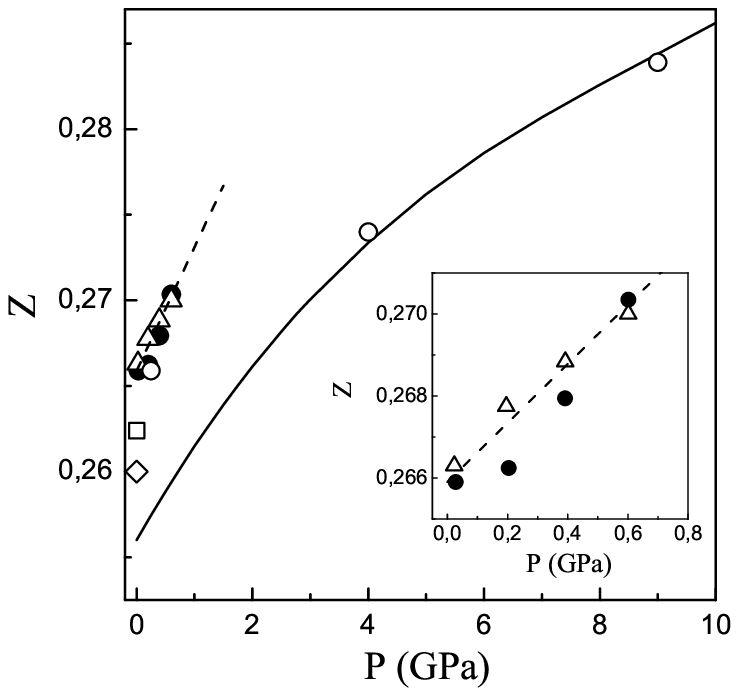}
\caption{\label{Z(P)} Pressure dependence of the internal
chalcogen structural parameter $Z$ for FeSe. The solid line
indicates the results of calculations from Ref.
\onlinecite{Ciechan}. Experimental data on parameter $Z$ in FeSe
for the tetragonal phase at $T=190$ K ($\bigtriangleup$, Ref.
\onlinecite{Millican}), $T=295$ K  ($\Box$, Ref.
\onlinecite{Gomez}), $T=300$ K ($\Diamond$, Ref. \onlinecite{Hu}),
and for the orthorhombic phase at $T=16$ K ({\Large$\circ$}, Ref.
\onlinecite{Margadonna}) and $T=50$ K ({\Large$\bullet$}, Ref.
\onlinecite{Millican}). In inset: data of Ref.
\onlinecite{Millican} on the expanded scale. }
\end{center}
\end{figure}

In order to clarify the behavior of $\chi (P)$ in FeSe, we carried
out the {\em ab initio} calculations of paramagnetic susceptibility
as a function of pressure.
These calculations are based upon the pressure dependent structural parameters,
which have been calculated and listed in Ref. \onlinecite{Ciechan}.
As is seen in Fig. \ref{Z(P)}, the calculated behavior of $Z(P)$ follows the
available experimental data \cite{Margadonna,Millican,Hu,Kumar,Gomez}.
In the course of corresponding calculations of $\chi (P)$ for FeSe we evaluated
the pressure derivative d\,ln$\chi$/d$P\simeq 8\times 10^{-2}$ GPa$^{-1}$
in the range 0$\div$1 GPa.
The evaluated derivative appeared to be in a qualitative agreement with
the experimental low temperature data (see Table \ref{dlnX/dP}).

\begin{figure}[]
\begin{center}
\includegraphics*[trim=0mm 0mm 0mm 0mm,scale=0.42]{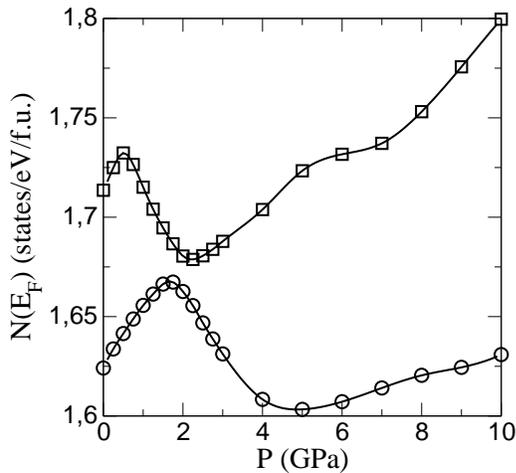}
\caption{\label{DOS(P)} Calculated pressure dependencies of the
density of states at the Fermi level for FeSe.
The pressure dependent structural parameters, including the
lattice constants and chalcogen atom position $Z$, were taken from
the calculations of Ref. \onlinecite{Ciechan} ({\Large$\circ$}).
For another set of $N(E_{\rm F})$ calculations ($\Box$) the small
upward shift $\Delta Z=+0.004$ was employed to start from the
experimental value $Z=0.26$ (see $Z(P)$ behavior in
Fig.~\ref{Z(P)}). The solid lines are guides for the eye.}
\end{center}
\end{figure}

As part of these electronic structure calculations for FeSe, we also obtained
a pressure dependence of the density of states at the Fermi level, which is
presented in Fig. \ref{DOS(P)}.
In addition to the calculated structural parameters from
Ref. \onlinecite{Ciechan}, we also employed the small upward shift
$\Delta Z=+0.004$ to start the optimized $Z(P)$ dependence from the
experimental value $Z=0.26$ (see $Z(P)$ behavior in Fig. \ref{Z(P)}).
The corresponding two sets of calculations demonstrate in Fig. \ref{DOS(P)}
a tolerable variation of $N(E_{\rm F})$ behaviors in FeSe under pressure,
depending on a small adjustment of $Z(P=0)$ between the theoretical and
experimental values.

With the aim to elucidate the main mechanism of the experimentally
observed strong increase of the magnetic susceptibility of
FeSe under pressure at low temperatures, we have also analyzed the pressure
effect in terms of the corresponding changes of the volume and
$Z$ parameters by using the relation:
\begin{eqnarray}
{{\rm d\,ln}\chi\over {\rm d}P}= {\partial\,{\rm ln}\chi\over
\partial\,{\rm ln}V}\times{{\rm d\,ln}V\over {\rm d}P}+{\partial\,{\rm ln}\chi\over
\partial Z}\times{{\rm d}Z\over {\rm d}P}.
\label{dX/dV}
\end{eqnarray}
The required values of the partial volume and $Z$ derivatives of
$\chi$ can be estimated from the results of {\em ab initio}
calculations presented in Figs. \ref{X(V)} and \ref{X(Z)}, and
were found to be $\partial\,{\rm ln}\chi/\partial\,{\rm ln}V\simeq
8$ and $\partial\,{\rm ln}\chi/\partial Z\simeq 65$ for the values
of $V$ and $Z$ close to the experimental data.
The optimized value d\,ln$V$/d$P=-3\times 10^{-2}$ GPa$^{-1}$
is taken for the compressibility of FeSe \cite{Ciechan}.
This calculated compressibility agrees closely with the experimental values
$-3.1\pm0.1 \times 10^{-2}$ GPa$^{-1}$~~\cite{Braithwaite,Millican,Garbarino}.
Also, the optimized value d$Z$/d$P\simeq 0.55\times 10^{-2}$ GPa$^{-1}$ was
adopted for evaluation of Eq. (\ref{dX/dV}).
As is seen in Fig. \ref{Z(P)}, this value of d$Z$/d$P$ is in agreement
with the experimental data at low pressures.
As far as all parameters entering Eq.~(\ref{dX/dV}) are estimated,
the first term in (\ref{dX/dV}) results in a large negative value
$\partial\,{\rm ln}\chi/\partial\,{\rm ln}V\times{\rm d\,ln}V/{\rm d}P\simeq
-24\times 10^{-2}$ GPa$^{-1}$, whereas the second term appears to
be large and positive: $\partial\,{\rm ln}\chi/\partial
Z\times{\rm d}Z/{\rm d}P\simeq 36\times 10^{-2}$ GPa$^{-1}$.
The both terms in Eq. (\ref{dX/dV}) taken together yield the estimate
d\,ln$\chi$/d$P\simeq 12\times 10^{-2}$ GPa$^{-1}$ for FeSe, which
is consistent with the low temperature experimental data.
This estimate is also close to the {\em ab initio} calculated pressure
derivative based on theoretical lattice parameters from Ref.
\onlinecite{Ciechan} (d\,ln$\chi$/d$P\simeq 8\times 10^{-2}$
GPa$^{-1}$, see Table \ref{dlnX/dP}).
Actually, the difference in
these evaluated values of d\,ln$\chi$/d$P$, depending on whether
experimental or theoretical lattice parameters are employed,
covers a reasonable range for the expected pressure effect on $\chi$.

Based on the results of calculations, presented in Figs.
\ref{X(V)} and \ref{X(Z)}, the observed hydrostatic pressure
effect on $\chi$ in FeSe at low temperatures can be represented as
a sum of two large in size and competing contributions, related to
the pressure dependence of the structural parameters $V$ and $Z$.
As a result, the experimental positive value of the pressure
effect, d\,ln$\chi$/d$P\simeq 10\times 10^{-2}$~GPa$^{-1}$, is
determined by a dominating contribution from the change of $Z$
under pressure.

Actually, the nature of the large positive pressure effect on
$\chi$ in FeSe is similar to that reported for FeTe compound
\cite{Grechnev0}.
However, in the case of FeTe such effect is
twice more pronounced, and also takes place at room temperatures,
whereas for FeSe d\,ln$\chi$/d$P$ is found to be negative at 300 K
(see  Table \ref{dlnX/dP}).
The reason of this difference is unclear and have to be elucidated.
At the present stage one can
presume, that the negative sign of d\,ln$\chi$/d$P$ derivative is
probably related to a nature of the observed anomalous growth of
$\chi (T)$ up to room temperatures (Fig.~\ref{X(T)}), which is not
the case for FeTe \cite{Grechnev0}.
It appears that at higher temperatures this anomalous growth of
$\chi (T)$ in FeSe is apparently reduced by applied pressure.

Basically, the observed positive pressure effect on $\chi$ in FeSe
at low temperatures correlates with the calculated increase of the
density of states at the Fermi level $N(E_{\rm F})$ at low
pressures (see Fig. \ref{DOS(P)}).
At higher pressures one can see nonmonotonic variation of $N(E_{\rm F})$
in Fig. \ref{DOS(P)} which clearly exhibits consecutive maximum and minimum.
It was recently shown \cite{Sadovskii}, that superconducting transition
temperatures $T_{\rm c}$ of a number of iron-based superconductors
correlate with the corresponding values of the density of states
at the Fermi level, thus supporting the BCS-like pairing mechanism
in these systems.
Remarkably, that the presently calculated
behavior of $N(E_{\rm F})$ under pressure (the upper curve in Fig.
\ref{DOS(P)}, with maximum at 0.5 GPa and minimum at 2.2 GPa) is
qualitatively consistent with the reported experimental
dependencies of $T_{\rm c}(P)$ in FeSe (corresponding maximum and
minimum of $T_{\rm c}(P)$ were observed at $P\simeq 0.8$ GPa and
$P\simeq 1.2$ GPa, respectively
\cite{Masaki,Miyoshi,BendelePRL,Bendele}).
The calculated pressure
dependence of DOS at the Fermi level for FeSe with the structural
parameters taken from Ref. \onlinecite{Ciechan} (the lower curve
in Fig. \ref{DOS(P)}) also contains consecutive maximum and
minimum of $N(E_{\rm F})$, which are substantially shifted to
higher pressures.

\section{Conclusions}
The magnetic susceptibility of FeSe compound is found to rise
substantially with temperature, which apparently points to the
itinerant nature of the electronic states of Fe.
The calculated paramagnetic susceptibility $\chi_{\rm para}(T)$ describes
qualitatively the experimental dependence $\chi(T)$, however the origin
of the observed about twofold increase of $\chi$ up to 300 K is puzzling.
From comparison of the experimental values of
susceptibility and its anisotropy with that calculated for
paramagnetic contributions to $\chi$ in FeSe, the additional
anisotropic diamagnetism is expected to be of the order of
$-1\times 10^{-4}$ emu/mol, which can relate to the diamagnetism
of conduction electrons and presumably has its origin in the
quasi-degenerate electronic states close to $E_{\rm F}$.

The measurements of magnetic susceptibility under hydrostatic
pressure revealed a strong positive effect at low temperatures.
This effect appeared to be comparable with
that reported for the related FeTe compound \cite{Grechnev0},
whereas at room temperature the pressure effect for FeSe is found
to be also strong, but {\em negative}.

Our calculations indicate that paramagnetic susceptibility of the
FeSe compound is substantially dependent on the unit cell volume
$V$ and the relative height $Z$ of Se species above the Fe plane.
It is shown that the observed at low temperatures large positive
pressure effect on $\chi$ is related to the strong sensitivity of
the paramagnetic susceptibility to the parameter $Z$,
which determines the dominant positive contribution.
The grounds of the negative sign of d\,ln$\chi$/d$P$ derivative
in FeSe at 300 K are unclear and probably linked to a nature
of the observed anomalous growth of $\chi (T)$.
At present one can state that at higher temperatures
this anomalous growth of $\chi (T)$ in FeSe is apparently reduced
by applied pressure.

The present {\em ab initio} calculations have demonstrated that
for FeSe compound the nonmonotonic behavior of superconducting
transition temperature with pressure qualitatively correlates with
the density of electronic states at the Fermi level.
This indicates a possible realization of the BCS-like pairing
mechanism in this system.

\begin{acknowledgments}
This work was supported by the Russian-Ukrainian RFBR-NASU project
01-02-12 and 12-02-90405, by Russian Ministry of Science and
Education through state contracts 11.519.11.6012 and
14.740.11.1365, by NASU Young Scientists Grant 03-2012, and by a
grant of the President of the Russian Federation for State Support
of Young Russian Scientists (MK-1557.2011.5).
\end{acknowledgments}

\end{document}